\documentclass[conference]{IEEEtran}
\IEEEoverridecommandlockouts

\usepackage[table,xcdraw]{xcolor}

\usepackage{multirow}
\usepackage{pdflscape}
\usepackage{amsmath,amssymb,amsfonts}
\usepackage{algorithmic}
\usepackage{graphicx}
\usepackage{textcomp}
\usepackage{longtable}
\usepackage{xcolor}
\usepackage{soul}
\usepackage{tabularx}
\usepackage{inputenc}
\def\BibTeX{{\rm B\kern-.05em{\sc i\kern-.025em b}\kern-.08em
    T\kern-.1667em\lower.7ex\hbox{E}\kern-.125emX}}
\usepackage[style = ieee]{biblatex}

\begin{document}

\title{Best Privacy Practice Recommendations for Global Audio Streaming Platforms \\}

\author{\IEEEauthorblockN{Annette Stawsky}
\IEEEauthorblockA{\textit{Carnegie Mellon University} \\
Pittsburgh, PA \\
astawsky@andrew.cmu.edu}
\and
\IEEEauthorblockN{Kang-Yu Wang}
\IEEEauthorblockA{\textit{Carnegie Mellon University} \\
Pittsburgh, PA \\
kangyuw@andrew.cmu.edu}
\and
\IEEEauthorblockN{Ye In Kim}
\IEEEauthorblockA{\textit{Spoon Radio} \\
Seoul, South Korea \\
grace@spoonradio.co}
\and
\IEEEauthorblockN{Dong Hyuk Shin}
\IEEEauthorblockA{\textit{Spoon Radio} \\
Seoul, South Korea \\
ethan@spoonradio.co}}

\maketitle
\thispagestyle{plain}
\pagestyle{plain}

\begin{abstract}
Spoon Radio is a rapidly growing global audio streaming platform which currently operates in South Korea, the United States, Japan as well as the Middle East and North Africa. The platform believes that its commitment to user privacy is an important competitive factor. As such, it aims to not just comply with existing privacy regulations in regions where it operates today but to also ensure that it anticipates likely evolution of these regulations and of user expectations. In doing so, Spoon Radio wants to ensure it is well prepared to continue its expansion into new markets.

As part of an effort to inform the evolution of its data practices, Spoon Radio reached out to the Privacy Engineering Program at CMU and sponsored a capstone project in which two master’s students in the Program worked with Spoon Radio personnel over the course of the 2021 Fall semester. The present report summarizes best practice recommendations that have emerged from this collaboration. These best practices are a combination of practices that are already implemented or in the process of being implemented by Spoon Radio today as well as more aspirational recommendations, which are expected to help inform Spoon Radio’s practices in the future.

In this report, best practice recommendations are organized around four stages of the data life cycle: data collection, data storage, data usage, and finally data destruction. A separate section is devoted to content moderation, an area where platforms such as Spoon Radio need to reconcile considerations such as promoting freedom of expression with the need to create a safe and respectful environment that complies with applicable laws and respects relevant cultural values.

\end{abstract}

\section{Introduction}
\subsection{Social Audio Streaming Platforms} Over the past decade, a number of sites have emerged that empower everyday users to host their own live streaming channels, both in audio and video formats. These sites provide users with tools to create content and share it with the world. YouTube is an example of a well-established platform that facilitates content creation. Snapchat, a more recent service than YouTube, introduced ephemeral videos. Users are encouraged to use Snapchat on an ongoing basis and send their friends quick glimpses into their everyday lives.

Increasingly, content creation platforms are integrating social features into their services. Providing direct message functionality for users to interact is one example of this trend. Spoon Radio is a platform that allows users to create audio streaming content and interact with one another. Spoon Radio combines features for creative content creation with features for social interaction to create a unique service centered around audio streaming. Unlike social platforms that look at audio streaming as an add-on to their existing functionality, Spoon Radio focuses on audio streaming and emphasizes user communication that relies primarily on audio streaming. In this report, we refer to this category of platforms as “social audio streaming platforms.”

\begin{figure}[htbp]
    \centerline{\includegraphics[scale = .5]{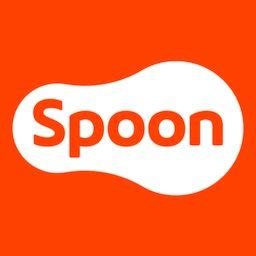}}
    \caption{Spoon Radio's Logo}
    \label{Spoon App}
\end{figure}

\subsection{Spoon Radio}
Spoon   Radio   is   a   rapidly   growing   global  audio streaming platform that currently operates in South Korea, the United States, Japan as well as the Middle East and North Africa. There are three main features in Spoon Radio’s service: “LIVE”, “CAST” and “TALK.” Each feature involves using audio streaming to interact in different social contexts.
\subsubsection{LIVE}
This is Spoon Radio's primary service in which a user hosts a streaming channel as the “streamer” and other users join and communicate with the host as “listeners.” The streamer can purchase additional live-stream functionality such as longer streaming hours. Listeners can comment and purchase gifts in the form of stickers as a public display of support during the streaming. Using the Live Call functionality, streamers and participants can communicate one-on-one or as a group. As the name suggests, live-streams are ephemeral to users unless the host stores them and publishes them using the ``CAST" feature.

\begin{figure*}[htbp]
    \centering
    \includegraphics[width = .65\textwidth]{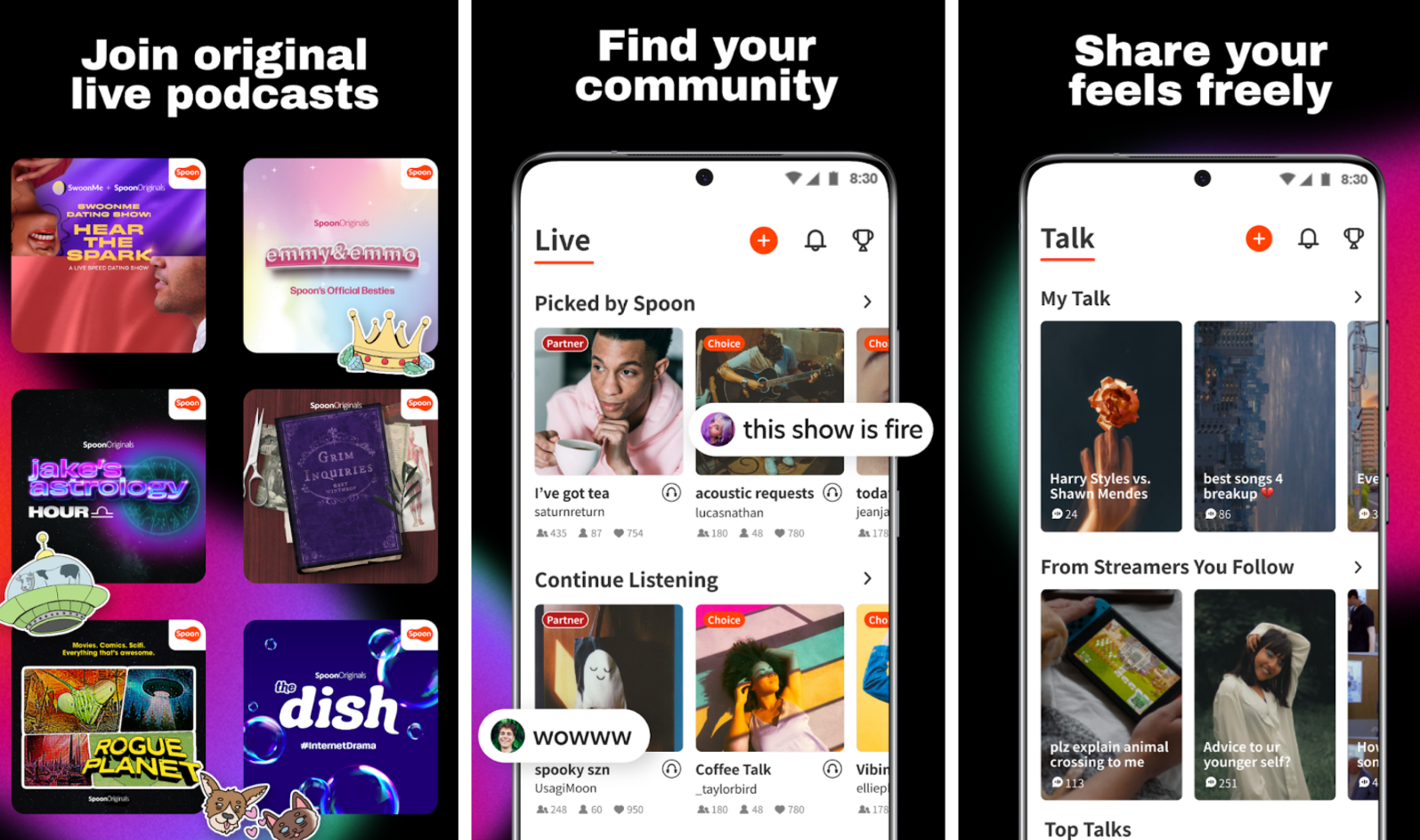}
    \caption{Spoon Radio's Three Services}
    \label{LIVE}
\end{figure*}

\subsubsection{CAST}
This feature allows the streamer to save and upload a previous “LIVE” session. The published ``CAST", short for podcast, remains on the streamer’s user profile. Any internet user can listen to these recordings, but only Spoon Radio users can comment and create recordings of their own. 

\subsubsection{TALK}
This feature allows users to create a personal chat room and communicate primarily through audio with each other, but the content remains public.

\subsection{Report Goals and Structure} 
As part of an effort to inform the evolution of its data practices, Spoon Radio reached out to the Privacy Engineering Program at CMU and sponsored a capstone project in which two master’s students in the Program worked with Spoon Radio personnel over the course of the 2021 Fall semester. The  present  report summarizes best  practice  recommendations  that  have  emerged from this  collaboration.  These  best  practices  are  a  combination of practices  that  are  already  implemented  or  in  the  process of being  implemented  by  Spoon  Radio  today  as  well  as  more aspirational recommendations, which are expected to help inform Spoon Radio’s  practices  in  the  future. 

The best practices compiled in this report are intended to be simple, practical recommendations to proactively protect user privacy. These best practices are based on privacy principles and academic research, and extend beyond minimal legal requirements. While recommendations provided in this report are intended for a global platform with users around the world, we pay particular attention to legal requirements in Japan, Korea and the US, as these markets are particularly important to Spoon Radio today. Where appropriate we also draw on other regulatory requirements such as those associated with GDPR \cite{GDPR_text}. When referencing Japan's Act on the Protection of Personal Information (APPI), we are referencing the update to APPI that comes into effect in 2022 \cite{OneTrust_APPI}.   

We organize best practices around 4 stages of the data life cycle: data collection, data storage, data usage, and data destruction. We dedicate a section to best practices for content moderation for social audio streaming platforms. Creating community rules that protect all users fairly is especially hard on a social audio streaming platform because of the challenges presented by user-generated content (e.g. content that may be factually wrong, content that may be offensive or defamatory, content that may express controversial views, content that may be inflammatory, etc.). Moderating such content requires balancing freedom of speech considerations while promoting a safe and respectful environment as well as complying with laws and cultural expectations that vary from one country to another. In addition, content moderation should be transparent, fair, consistent and efficient.

One common definition of privacy revolves around `the right to be let alone,' `free from undue interference from others' \cite{letalone, brandeis1890right}. Content moderation has to have a sufficiently light touch to not infringe on the right of people to express their opinions, yet it has to also be sufficiently effective to protect people from undue interference from libel, from defamatory speech or other forms of expression that would interfere with their legitimate expectations of safety and respect.

How far should these responsibilities be taken? Should moderation extend to taking down content that may be revealing sensitive information about others without their consent such as a streamer unintentionally revealing something private about someone else without their consent (e.g., a health ailment, an addiction, a particular sexual orientation). Should the platform have mechanisms to remind streamers to carefully think about the content they generate? Should it have mechanisms for someone whose private life has been exposed or who feel offended by content posted via the platform to request that the content be taken down? These and other examples are discussed in this report along with some initial recommendations. The report also acknowledges that issues such as moderation are particularly complex and will require further analysis and refinement over time, as new lessons are learned and as social norms and regulatory expectations continue to evolve.


\section{Background} 
In this section, we give a brief history of Fair Information Practice Principles (FIPPs) and justify using them as a basis for our recommendations \cite{fipps}. We further discuss how user-generated content, such as audio recordings and comments, present  unique privacy challenges. 

\subsection{Fair Information Practice Principles}
\begin{figure*}[htbp]
    \centerline{\includegraphics[width=\textwidth]{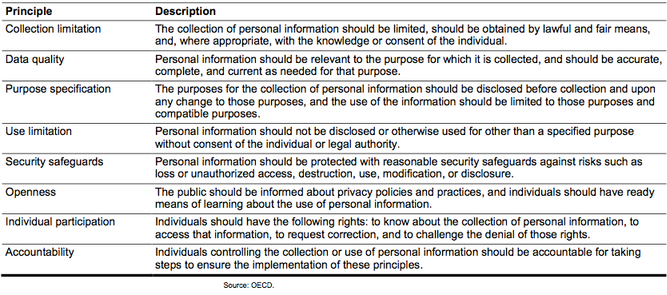}}
    \caption{The OECD's 8 Fair Information Practice Principles \cite{oecd}}
    \label{OECDFIP}
\end{figure*}

We base our best practice recommendations in part on aspirational interpretations of fair information practice principles (FIPPs), including specific regulations, as well as on academic research in privacy. FIPPs have evolved over the past 50 years, beginning in 1973 when the United States Department of Health, Education and Welfare published the `HEW Report' outlining 5 foundational principles \cite{hew}. In 1980, the Organization of Economic Cooperation and Development (OECD) built upon the HEW Report and introduced the 8 FIPPs that we will reference in this report \cite{oecd}. These 8 principles are detailed in Figure \ref{OECDFIP}.

FIPPs have influenced privacy regulations around the world, including in the EU (GDPR Art. 5), U.S., Canada, Sweden, Australia, and Belgium to name just a few \cite{GDPR_text, hew, historyfipps}. Interpretations of FIPPs vary from one jurisdiction to another and also continue to evolve with regulations introducing increasingly more stringent requirements and penalties. Taking an aspirational view of FIPPS informed by recommendations coming out of research enables us to develop a set of recommendations that we believe are more likely to stand the test of time. These recommendations, we hope, will position Spoon Radio and other global audio streaming platforms that adopt them for continued expansion into new jurisdictions and prepare them for the way in which current regulation is likely to evolve.  

The notice-and-choice framework encapsulated in FIPPS has been criticized for being impractical, in part because of  usability challenges such as people not having the time to read privacy policies or exercise choices offered to them \cite{cate2006failure}. Recommendations made in this report take this criticism into account and include suggestions based on research aimed at overcoming these limitations with a particular focus on usability issues.

\subsection{Challenges Associated with the Collection of User-Generated Content}
Social audio streaming platforms present unique privacy challenges because they necessarily collect user-generated content, including audio recordings. In this section, we define structured personal data and unstructured, user-generated content. We analyze the difficulties associated with user-generated content in Table \ref{freeform} below. We use the standard terminology defined in GDPR, Article 4 for ``Personal Data," ``Data Controller," and ``Data Subject" \cite{GDPR_text}.
\subsubsection{Structured Personal Data}“Structured personal data” includes any personal data where the content is specific and known or predetermined by the data controller. An example of structured personal data is an online form where a user can enter specific personal information such as an email address. In this example, the data controller knows the exact data elements they are receiving and how sensitive they are.

If a user is prompted to list their personal interests to inform recommendations on a platform, this can be done in a structured way or in an unstructured/free-form way. Providing the user with a list of topics to pick from would make these interest structured personal data, whereas allowing for free response would make it user-generated content. 

\subsubsection{User-Generated Content}“User-Generated Content (or Unstructured Content)” includes any content where the user has control over which personal data elements, if any, to include. Audio recordings/live streams and comments are examples of unstructured content.

This report covers a wide range of best practice recommendations, some of which are widely adopted already and well-understood, while others are more challenging to implement. In general, best practice recommendations associated with free-form, user-generated content tend to be more difficult to articulate and implement than those associated with structured data. Table \ref{freeform} below compares the difficulty of implementing privacy objectives in structured data as opposed to user-generated content. As an illustration, the table considers 4 particular privacy objectives: providing effective privacy notices to users, moderating content in a fair, effective and transparent fashion, securing sensitive data, and responding to subject access  requests (e.g. user requests to access or delete their data).

These four objectives are selected because they represent themes that show up repeatedly throughout the rest of the report: notices, moderation, security, and user rights. We use a three point scale for this comparison: 1 = Easily understood and readily implementable, 2 = Generally well understood and amenable to implementation subject to some minor adaptation, 3 = not fully understood with implementation requiring a high degree of sophistication and some level of innovation.

\begin{table*}[htbp]
\centering
\caption{Difficulty analysis for defining and implementing mechanisms to support privacy objectives in the context of structured data and free-form user generated content. \\ A 3 point scale is used to provide an estimate of the difficulty of implementing each recommendation – with 1 being easy and 3 being the most challenging.}
\begin{tabular}{|p{0.25\linewidth}|p{0.30\linewidth}|p{0.30\linewidth}|}
\hline
\textbf{Privacy Objectives} & \textbf{Structured Data} & \textbf{User-Generated Content (e.g. Audio Streaming, Comments, etc.)}\\
\hline 
Provide adequate privacy \textbf{notices} & \textbf{Score: 1.} 

Structured data implies that data types and their possible values are generally known at design time. Accordingly privacy notices can explicitly detail their practices as they pertain to the collection and use of each data type. Developing such notices is simply a matter of knowing each type of data being collected and how that data is handled (e.g. for how long it is stored, for what purpose, with whom it is shared, etc.). This being said, the platform should aim to make notices as visible and effective as possible. & \textbf{Score: 1. }

Practices pertaining to the collection, use, sharing, retention, etc of user-generated content can easily be disclosed in a privacy notice. This being said, the platform should aim to make notices as visible and effective as possible.  \\
\hline
Establish clear and consistent \textbf{moderating rules} & \textbf{Not applicable}, except possibly for a limited set of situations (e.g. user adopting an offensive user name or filling public elements of his or her public profile with inappropriate content)
 & \textbf{Score: 3. }
 
Establishing clear, fair, and effective moderation principles is challenging. In addition, implementing such principles is difficult to automate and may require significant human resources.

\\
\hline
Enhance security \textbf{safeguards for sensitive data} & \textbf{Score: 1.} 

Standard security best practices can be used to safeguard the security of sensitive data (e.g., encryption of sensitive data, least privilege access control, etc.)
& \textbf{Score: 2.} 

Securing user-generated content may require distinguishing between \textbf{public and private content}, the latter for instance in the content of private streams only accessible to a subset of users. Access control requirements to control not just who has access to the content but possibly also who is allowed to generate or possibly even modify some content (e.g. if multiple people contribute to some content) can possibly become more complex. Providing users with usable controls to determine who can access their content is not entirely trivial. \\
\hline
Respond to user requests for deletion, access, correction, etc. & \textbf{Score: 1.} 

Mechanisms to support user requests for deletion, access to, or correction of their data are generally well understood and straightforward to implement & \textbf{Score: 3.} 

Because user-generated content may include information about people other than the content creator (as well as other sensitive topics), issues of deletion, access and correction are more complex than for structured data about the user himself/herself. Some of these issues fall under the broader topic of content moderation (see above).

\\
\hline
\end{tabular}
\label{freeform}
\end{table*}

It is true that most organizations collect some user-generated content if they allow data subjects to call, email, or submit a ticket to customer support. However, there are three components that distinguish a social audio streaming platform’s user-generated content from standard user-generated content such as customer support.
\begin{enumerate}
    \item Increased visibility: The primary functionality is to create a user profile full of audio recordings. These recordings and comments on these recordings are visible to the public as opposed to just customer support staff.
    \item Increased frequency: A user provides the audio streaming platform with user-generated content every time they use the main functionality of the service, as opposed to whenever they have a complaint or technical problem. 
    \item Increased likelihood of the content being sensitive: The nature of user-generated content on an audio streaming platform is far more likely to be sensitive or emotional because users are expressing themselves creatively and trying to form connections with each other.
\end{enumerate}

The wide variety of topics covered in audio streaming content further complicates its analysis and moderation, as it is impossible to anticipate all possible discussions and scenarios. Throughout this report, we discuss best practices within the context of the challenges mentioned in this section.

\section{Best Practices for Data Collection}

Data collection is the first stage of the data life-cycle in which a social audio streaming platform gathers personal data provided by the data subject themselves, by monitoring or tracking the data subject, or from other third parties. In line with a bare minimum application of the openness principle, a social audio streaming platform's data collection practices should be detailed clearly and specifically in a privacy notice that is provided to users at or before the point/time of collection \cite{10.1007/978-3-642-03168-7_3}. 

Beyond this minimum recommendation, we recommend adopting disclosure mechanisms that can be expected to further enhance user awareness of the platform’s data practices. This includes the development of more prominent, context-sensitive notices that are presented to users in a timely fashion such as notifying them about data collection as it is about to take place or providing them with dashboards that feature summaries of the data that has been collected about them and give users ready access to controls where they can exercise relevant rights about the retention and use of that data. 

\begin{figure*}[htbp]
    \centerline{\includegraphics[width=\textwidth]{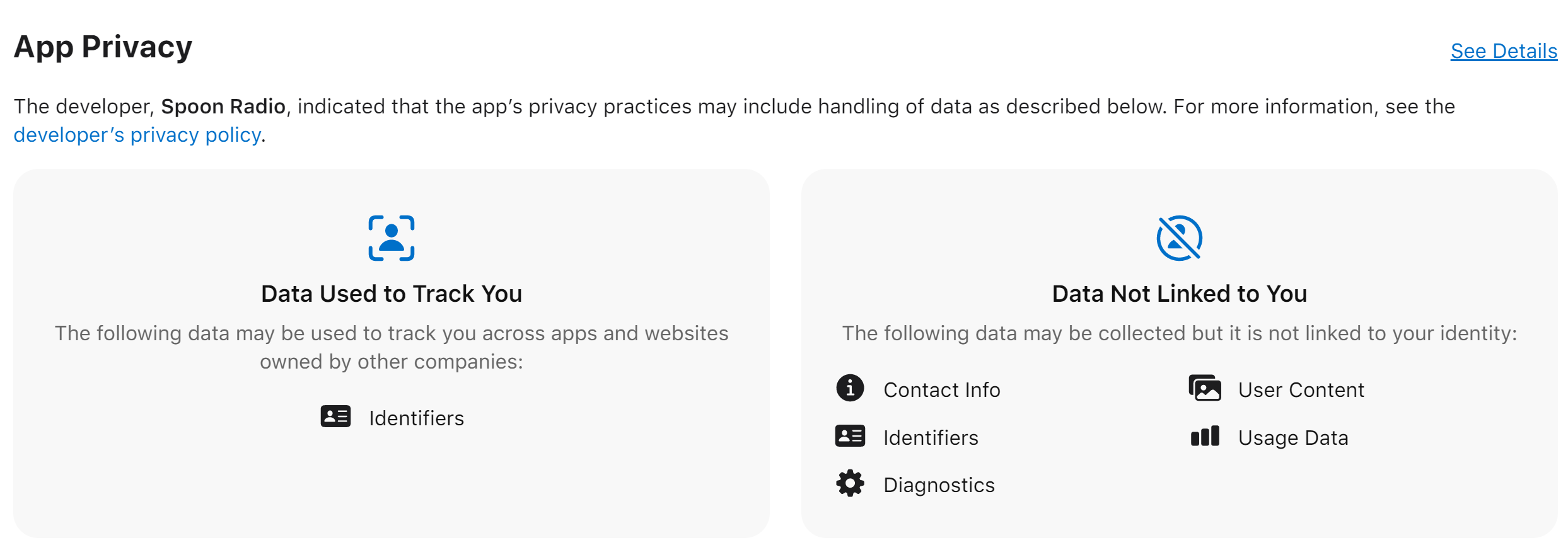}}
    \caption{Spoon Radio's Privacy Nutrition Label on Apple's App Store \cite{spoonapplestore}}
    \label{nutritionlabel}
\end{figure*}

Privacy nutrition labels are usable notices that summarize salient details from a complex privacy policy \cite{kelley2010standardizing}. Users could also be provided with simple settings that enable them to customize those elements of nutrition labels they care to be notified about \cite{kelley2010standardizing}. It is unrealistic to expect users to read and understand every privacy policy they encounter- one study estimated the time an individual would have to spend reading the policies of every site they access at 244 hours a year, with an estimated opportunity cost of $\$3,534$ a year \cite{mcdonald2008cost}. That amounts to a national $\$781$ billion opportunity cost (for the United States population) to read privacy policies that are often vague and ambiguous \cite{mcdonald2008cost, anthonysamy2011privacy}. It follows that privacy nutrition labels could do far more than save the user time. They could provide the user with information that is otherwise impractical to obtain and allow the user to make better informed privacy decisions. Apple requires new apps to fill out a privacy nutrition label for display in the app store \cite{nutlabel_apple}. Figure \ref{nutritionlabel} shows Spoon Radio's privacy nutrition label on the Apple app store. 

In privacy, information asymmetry refers to the significant disparity between the information known by an organization and that known by a user. In the absence of usable notices or additional privacy regulations, current privacy notices do not solve problems such as information asymmetry \cite{bashir2015online}. 
We recommend adopting best practices that are practical for the current legal landscape while recognizing that further improvements, such as standardizing privacy policies, are necessary for meaningful privacy controls.

In line with the collection limitation principle, a social audio streaming platform should not collect more data than is strictly necessary to comply with the specified purposes of collection \cite{10.1007/978-3-642-03168-7_3}.
Date of Birth is an example of a data element that is commonly collected, but not always strictly necessary. If the primary purpose for collecting a user’s date of birth is to enforce age requirements, then collecting the user’s birth year and month could be sufficient. Some social networks offer users discounts on their birthdays, or publicize birthdays to other users. Both of these objectives could be accomplished using the user’s birthday month or birthday week instead. Social audio streaming platforms should consider collecting more granular data elements whenever it can be reconciled with the purpose of the data. 

In order to provide an informative notice, a social audio streaming platform should carefully consider the purposes for which each personal data element is collected and the security measures in place.

\subsection{Purpose Specification}
It is crucial to include accurate, comprehensive purpose specifications in a notice to allow users to make an informed decision about their personal data \cite{reidenberg2015disagreeable}. In line with the purpose specification principle, an adequate notice should clearly and specifically link each data element to the purpose of collection. Table \ref{purposespecification} displays reasonable purposes of collection for social audio streaming platforms, how the purposes relate to specific data elements, and how the purposes could be made more specific and informative. For a social audio streaming platform, the example purposes that we will refine and improve are: 
\begin{enumerate}
    \item data is used/necessary for moderating purposes
    \item data is used/necessary to train employees on data protection
    \item data is used/necessary to comply with legal requirements
    \item data is used/necessary to provide the standard service 
    \item data is used/necessary to improve and enhance services
    \item data is used/necessary to communicate with the user
\end{enumerate}

Evidently, social audio streaming platforms can have a wide range of valid purposes for collecting personal data, but all of these purposes should be phrased in clear, specific language and limited in scope. Purposes that are reasonable but not strictly necessary should be presented as a choice to the user, ideally an opt-in choice rather than opt-out. Opt-out choices are notoriously difficult for users to find and become ineffective choice mechanisms \cite{JohnnyOptOut}. Additionally, many users are not motivated to go searching for opt-out choices and many companies are not motivates to inform users of their choices in meaningful ways. If an opt-out mechanism is used, it should be easy to discover and use, requiring no more than a few clicks from the user's homepage \cite{JohnnyOptOut}. We recommend opt-in choices to encourage transparency before data collection takes place. 

As shown in Table \ref{purposespecification}, we propose identifying specific scenarios as an effective way of limiting the scope of a purpose. Furthermore, we propose answering questions that a potential user might have about the purpose as an effective way of refining the language of the purpose. Table \ref{purposespecification} illustrates how social audio streaming platforms can use these two tactics to refine the language and scope of the relevant purposes listed above. 

The process of linking each data element to its purpose(s) is iterative. Throughout the process of mapping each data element to its purpose, social audio streaming platforms will be able to identify unnecessary data and cease collection. Then the mapping should be verified with relevant teams within a social audio streaming platform and then included in the public privacy policy (and other notices such as nutrition labels, as appropriate). For social audio streaming platforms, the mapping should be verified with the team that moderates content, marketing, software engineering, and teams that are responsible for tasks such as profiling a user, integrating with third parties, and storing personal data.

Minimally, a privacy notice should include the date that it was last updated, as well as a table of edits made to the notice, and links to previous versions of the notice.  Ultimately, best practices recommend that a company create an exhaustive list of purposes and abide by it. In practice, this level of foresight is difficult in any organization and strategic decisions could require that changes be made to the purpose list. If a social audio streaming platform identifies a need to change the purpose list, the notice itself should be edited and data subjects should be notified (if the change only applies to personal data) or applied to for consent (if the change applies to sensitive data). APPI, CPRA, and GDPR require notifying users of changes to privacy notices.

Occasionally, organizations include language in their notices saying that anonymized data can be used for any purpose. Many organizations have made the mistake of inadequately anonymizing data, publishing it, and harming their users' privacy \cite{porter2008identified} \cite{samarati1998generalizing}. Social audio streaming platforms should establish effective procedures for anonymization and adequate training for employees involved. Doing this type of analysis in a thorough manner involves looking for all possible data-sets available that might lead to de-anonymization before claiming that the data set is anonymous. This includes public data sets, data sets that private organizations have access to, and data sets these organizations might share with other third parties \cite{sweeney2000simple}.

\subsection{Security Measures} 
An adequate notice should contain far more than a list of data elements collected. In line with the openness principle, a notice should inform users about what data is collected and for what purpose. 

Audio content and comment content are user-generated and could contain sensitive information if the user chooses to include it. Though users create audio content directly on the service, a social audio streaming platform might offer the functionality to integrate their profile with other social media profiles so that the user only has to manage one centralized profile. If this functionality is offered, a social audio streaming platform should notify the user that any sensitive content they created on the third party profiles will now be collected and processed/stored according to their privacy standards. In line with the Data Quality principle, social audio streaming platforms should verify the identity of the third party providing the personal data. In section IV we will further discuss the rights that best practices recommend for data subjects.


The mechanism of transferring and storing personal data is crucial to ensure that data is protected in transit. Taking reasonable measures to protect the security of personal data in transit and at rest is not only part of FIPPs, but also a requirement found in most privacy regulations (CPRA and APPI included).

\begin{table*}[!ht] \setlength\tabcolsep{5pt} 
\caption{Purpose Specification Table for a Social Audio Streaming Platform}
\begin{center}
\begin{tabular}{|p{0.12\linewidth} | p{0.22\linewidth}| p{0.3\linewidth}|p{0.3\linewidth}|} 
\hline
\textbf{Purpose} & \textbf{Applicable Data Elements} & \textbf{Refined Language} & \textbf{Refined Scope}\\
\hline
Data is used/necessary for \textbf{moderating} purposes & Full Name, Nickname/ Username, Content of audio recordings and live streams, Content of comments, Contents of Communications with the Social Audio Streaming Platform (emails or calls to Customer Service) & \textbf{Why is moderating done?} Audio and comment content is moderated to protect user privacy and to ensure users are abiding by community guidelines.

\textbf{How is moderating done?} There is a team of trained employees responsible for moderation of content posted to this platform. Additionally, there is a machine learning model that will conduct preliminary reviews of all audio and comment data, and users are encouraged to report violations to community guidelines.

\textbf{Are there consequences for repeated offenders?} Yes, the privacy policy should note that data about violations (user, date, content containing the violation, etc.) is retained and used in future moderating cases. Repeated offenders are subject to stronger penalties including account suspension and termination. & There are three scenarios where moderating applies.
\begin{enumerate}
    \item all audio and comment data undergoes a preliminary moderating process that is automatic and checks for violations to the community rules of conduct. Unflagged content will not be subject to additional moderation unless any of the following two scenarios apply. Flagged content will be transferred to the moderating team and handled in accordance with the community guidelines procedure.
    \item relevant data that was reported by another user will be sent to the ML model to flag violations AND sent to the moderation team for review. 
    \item complaints or reports filed by another user where the data in question includes personal information about a third party user. This scenario will be handled in the same way as the second scenario above with the addition that the third party user and the user who filed the complaint will be notified of the moderating decision.
\end{enumerate}
\\
\hline
Data is used/necessary to \textbf{train employees} & Content of audio recordings and live streams, Content of comments, Contents of Communications with a social audio streaming platform (emails or calls to Customer Service) & \textbf{Will all employees be trained using this data?}
The only employees that will be trained using the relevant personal data are moderators before they join the moderating team. 

\textbf{Will the data be protected before use in this way?}
The content of the audio, comment, or complaint will be left intact, but the surrounding metadata will be scrubbed of any personally identifiable data. For example, if a comment is used, the name of the user will not be included, unless the user themselves mentions it as part of the comment content.
 & As a moderating team handles a specific scenario, the manager can choose to flag the scenario for use in training. Flagged scenarios are then pseudonymized by removing names and other identifiers not included in the content itself. The pseudonymization is automated using a program, and only a subset of content pseudonymized is actually used in training. 
 
The manager’s experience and interactions with the team qualify him/her to identify scenarios that will be useful in training. However, the pseudonymized data might also be viewed by HR or talent acquisition teams that are responsible for compiling the training itself.
\\
\hline
Data is used/necessary to comply with \textbf{legal and judicial requirements} & All personal data elements listed above could potentially be subject to this purpose & \textbf{Where do these  requirements originate?}
Local police departments and government agencies are examples of potential parties involved in this scenario. 
& This is an example of a purpose where it may be imprudent to attempt an exhaustive list of scenarios. Since an organization may be obliged to provide personal data if requested by government officials, for example, attempting an exhaustive scenario list might mislead a user. 

However, the organization should indicate if it has a response plan for challenging warrants or whether it does not believe itself to be subject to these requests by federal officials. This analysis depends on the country the data is processed/stored and the user population. The organization should include a link to their transparency page.

Some example scenarios, though not an exhaustive list, include a police investigation into a user of the service, and a warrant provided by a government agency.
\\
\hline
\end{tabular}
\end{center}
\label{purposespecification}
\end{table*}

\begin{table*}[!ht] \setlength\tabcolsep{5pt} 
\caption{Continued Purpose Specification Table for a Social Audio Streaming Platform}
\begin{center}
\begin{tabular}{|p{0.12\linewidth} | p{0.22\linewidth}| p{0.3\linewidth}|p{0.3\linewidth}|} 
\hline
\textbf{Purpose} & \textbf{Applicable Data Elements} & \textbf{Refined Language} & \textbf{Refined Scope}\\
\hline
Data is used/necessary to \textbf{provide the standard service} & All personal data elements listed above could potentially be subject to this purpose (excluding financial information and third party cookies) & \textbf{Are all of the data elements required for every element of functionality of the service?} No. Some data elements are required just to create an account, but others are only required if a user chooses to make use of a specific feature. & The following data elements are used in the following scenarios: 
    \begin{enumerate}
        \item Required to create an account: Full Name, Password, Email Address, Birth Year (to confirm that the user does not violate age requirements), Identification (to verify user identity), IP Address (to confirm the country and regulations that apply to the user), Device and Software Information (to display the right design for a mobile application vs a computer), first party cookies or tracking technology (to keep the user logged in, etc.) 
        \item Optional to create a profile: Voice Introduction, Profile Photo, Gender, Topics of Interest (to receive better recommendations for users and content to follow), Demographic Information received from third parties (if a user signs in through Facebook, Gmail, etc.)
        \item For an audio livestream/recording (such as Spoon’s LIVE, CAST, or TALK services): audio livestreams/recordings and potential comments/photos included in comments
        \item To buy additional features (such as Spoon’s stickers): Financial information and payment metadata
    \end{enumerate}
    \\
\hline
    Data is used/necessary to \textbf{improve and enhance services} &
    \begin{itemize}
        \item Interaction Details (pages viewed, language selection, searches conducted, people followed, etc.)
        \item Demographic information received from third parties if a user logs in through a third party (e.g. Facebook, Gmail, etc.)
        \item Device and Software Information (web browser, operating system, phone carrier)
        \item Contents of Communications with Spoon Radio (emails or calls to Customer Service)
        \item Gender
        \item Birth Year
        \item Topics of Interest
    \end{itemize} 
    & \textbf{Are third parties involved in this process? } Many organizations use third parties for analytics. Any third parties used should be mentioned in this purpose as well, as the data that will be shared with them, and a link to the third party processor list to learn more about them. 
    & There are three scenarios where personal data might be used to improve and enhance services: 
    In all of these scenarios, demographic information, gender, topics of interest, and birth year will be used to analyze if trends vary according to these factors. Device and Software information will be used if a service is specific to mobile applications or a specific operating system. 
    \begin{enumerate}
        \item Remove unpopular features or services: Interaction Details are used to analyze which types of services are most liked and used. The platform can regularly review activity on services and remove services that do not perform well. 
        \item Develop user studies: Feedback received from emails will be occasionally used as prompts for user studies and surveys to analyze how prevalent concerns are and potential fixes. Users may volunteer to be part of a study beta testing new features. Certain studies could include any information necessary to provide the service and the new feature. The user will be informed of the purpose and procedure of the study before consenting to participate. Any sensitive data that is not necessary for this purpose will be scrubbed before use. 
        \item Training moderators could be seen as improving services. It was described in detail earlier in this table.
    \end{enumerate}
    \\
    \hline
    \end{tabular}
    \end{center}
\end{table*}
\begin{table*}[!ht] \setlength\tabcolsep{5pt} 
\caption{Continued Purpose Specification Table for a Social Audio Streaming Platform}
\begin{center}
\begin{tabular}{|p{0.12\linewidth} | p{0.22\linewidth}| p{0.3\linewidth}|p{0.3\linewidth}|} 
\hline
\textbf{Purpose} & \textbf{Applicable Data Elements} & \textbf{Refined Language} & \textbf{Refined Scope}\\
\hline
Data is used/necessary to communicate with the user & 
\begin{itemize}
    \item Full Name, Nickname/ Username
    \item Email Address
    \item Phone numbers
    \item Contents of Communications with Spoon Radio (emails or calls to Customer Service)
\end{itemize}
& \textbf{How can the organization be sure it is communicating with the right user?}
The user’s identity will be verified when the user creates an account (either by asking the user to confirm their email or to provide further identification). Only verified emails will be used for informing users of updates and data breaches.
& The following are 6 illustrative scenarios where the organization might need to communicate with the user:
\begin{enumerate}
    \item Inform users of updates to services or notices.
    \item Inform users of a data breach. 
    \item Verify the account upon creation. 
    \item Respond to questions or complaints filed to customer support.
    \item Update user A on the result of moderation in a case where user B included A’s personal information in B’s content.
    \item Send mobile push notifications for content that a user might be interested in, if the user is using the mobile application.
\end{enumerate}
\\
\hline
\end{tabular}
\end{center}
\end{table*}
\section{Best Practices for Data Storage}
Data storage is the second stage of the data life-cycle in which social audio streaming platforms are responsible for the particular way in which it stores information it collects and for how long. It also covers all security aspects relating to the storage of sensitive data. 
Following well-established data minimization principles, data should be stored at the level of detail at which it is actually needed and should only be kept for as long as needed \cite{fipps}. As detailed below, a social audio streaming platform should sanitize data whenever possible. This may include anonymizing data, aggregating data or possibly adding noise to the data that has been collected.

Many social networking sites include features that allow users to control who can see their posts. On Youtube, a video can be made public, private, or accessible only by a restricted group of people \cite{youtube_settings}. On Facebook, photos can be made public, private, visible to friends, or visible to a select group of friends \cite{fb_settings}. In line with Nissenbaum's contextual approach to privacy, additional protective measures should be implemented for personal data that a user considers to be more private (e.g. recordings that the user categorized as 'private') \cite{nissenbaum}. Users should be provided with privacy controls that enable them to adequately restrict access to their data. Such controls have to be sufficient to accommodate the diverse privacy preferences of members of the community. For a platform such as Spoon Radio that provides an ephemeral audio service (``LIVE"), users should be notified that these recordings are only ephemeral to other users and that the platform itself stores them. 

\subsection{Pseudonymization}
Pseudonymization is the process of replacing identifying data with arbitrary identifiers in order to distance a user's data from their identity. Pseudonymization does not guarantee anonymity and should be used in conjunction with other security and privacy practices. Audio recordings that are part of the user’s profile (such as audio recordings from Spoon Radio’s ``CAST" or ``TALK" service) must be directly linked to a user in order to provide the standard service. However, audio recordings that are ephemeral to users and that the creator does not wish to access anymore should not be directly linked to the user. 

Pseudonymization can be done using logical or physical isolation where perhaps the only link between an audio live-stream and the user that created it is a random user identifier. The mapping between a user and his or her pseudonym should ideally be kept on another server than the one where the pseudonymized content is stored. This way, if the server with the pseudonymized content is compromised, it will be harder for an attacker to re-identify the creators of the live streams. Though this attacker might be able to listen to each recording, use other sources of data, and possibly re-identify the creator, such an attack would likely require a significant amount of manual effort and creativity and would not scale to thousands of users. Additionally, if an attacker wishes to gather all ephemeral audio recordings for a specific user, this attack of iterating through the entire database to deanonymize each recording is extremely inefficient. 

Users are creating profiles on social audio streaming platforms in which they are being creative, expressing themselves, and relying on the duration of these audios. Spoon Radio, for example, considers audio content to be the intellectual property of the user that creates it. Users might be using their profiles as portfolios in support of job applications, or they might simply want to keep a repository of the content they create for future possible access. Data security is multifaceted; sometimes anonymity is the focus, but sometimes data quality and retrievability are the focus. Due to the attachment that users might have to their audio recordings, ensuring retrievability of these recordings should be a security priority. 

If applicable, it is recommended to keep a log of changes made over time, allowing the user to navigate back to earlier versions. One way to ensure that the audio content users create are retrievable is to establish an efficient backup system to protect against data tampering or accidental deletion. We recommend that social audio streaming platforms implementing access control mechanisms where permission to modify the content is limited to the creator.

This is another example in which ephemeral audio content might be subject to different recommendations. In particular, since users are not relying on accessing this content in the future and they are only stored for moderation, they should be removed as soon as that process is complete if no complaints/reports were filed and if the automatic moderation screening did not flag the content. This approach includes the risk of users evading consequences by publishing inappropriate content ephemerally. Social audio streaming platforms should weigh this risk with other considerations (such as the popularity of their ephemeral service and available moderating solutions) in order to find a balance in line with their organizational objectives.

\subsection{Storage Format} 
A social audio streaming platform should be prepared to provide a data subject with a copy of all their personal data, and provide it to a third party upon the data subject’s request. These are two separate requests: requests to review one's data, and portability requests. It is recommended that social audio streaming platforms create an automated mechanism to easily transfer the data in machine and human readable formats. Audio recordings are naturally machine readable. The organization and labeling of the audio recordings should also be easily understood by the average user to satisfy the first category of requests, requests to view one's data. Privacy regulations around the world often impose legal requirements for how soon an organization should respond to these two user requests. Having a automated mechanisms for transferring data allows social audio streaming platforms to reliably comply with these time constraints. 

When organizations respond to access requests, they often provide users with a compressed file that the user can download onto their local device to access the personal data. Since audio data is expensive to store, one practical solution that could enable social audio streaming platforms to comply with this best practice is to offer the compressed file for a limited amount of time. Multiple email reminders should be sent to the data subject during that period to remind them to download their file before it is deleted. Facebook is using this tactic, they only offer the compressed file for a few days before it is deleted \cite{fbAccessData}. Though it may seem that this limits the data subject’s right of access, a user can always submit another request. 

\subsection{Responding to Subject Access Requests}

Though privacy rights may vary depending on where the user lives and what regulations protect them, the individual participation and openness principles support consistent practices for all users regardless of relevant regulations \cite{fipps}. This can be more computational efficient for the social audio streaming platform because it removes the overhead of monitoring where users live and which rights apply to them. Furthermore, it can be easier to create notices if there are clear guidelines that always apply. Not only is this a best practice from a privacy perspective, but it can also establish trust between the user and the social audio streaming platform. Such practices can be advertised as a strategic advantage to potential customers. 

In line with the openness principle and building trust with consumers, a social audio streaming platform could remind users of their rights. One way of achieving this is to have a button on the homepage that prompts the user to review the data collected about them by accessing their dashboard.  The dashboard itself could provide them with ways of accessing their data, request modifications, etc and could also include a summary of pending requests such as deletion or portability requests, or complaints they have filed or others have filed about their content. 

This dashboard is a public way of expressing that the social audio streaming platform takes user rights and privacy seriously. It also publicly encourages users to take advantage of their privacy rights without discrimination. Though privacy regulations often set an upper bound on response time, this dashboard could also display the average response time for various requests so users can manage expectations.  
\section{Best Practices for Data Usage}

Information flows in social audio streaming plat- forms can be complex and involve multiple parties. A piece of collected data might be used and shared by multiple employees and automated programs of the platform (First Party), and Third Party services that are deeply integrated in the platform. For instance, when the user starts a streaming event, they first upload the audio information to the platform. This piece of data might go through several servers. And then the audience is able to listen to it. During this process, the platform moderators (First Party) may need to access the stream to check whether the content confirms with relevant platform guidelines. There may also be Third Party services that collect user behavior data at the same time. An important part of addressing privacy requirements when it comes to Data Usage has to do with disclosing and limiting the purpose(s) for which data is used, including who that data might be shared with and for what purpose.

In the following section, we first discuss issues that relate to the control users should be given over the usage of their data - user controls. This includes looking at data subject rights, adopting privacy dashboards, and notice and consent mechanisms. Then we discuss the best practice that considers the internal use of data: Policy, Access Control, and Third Parties.

\subsection{User Notification and Control}

\paragraph{Data Subject Rights}

Recent privacy regulations such as CCPA/CPRA or GDPR are increasingly mandating a number of data subject rights designed to empower data subjects to review their data, request that it be deleted, restrict the way in which it can be used and more. We now discuss the following relevant Data Subject Rights \cite{privacyperfect2021Data, truevault2021What, california2018cpra, gdpr2018gdpr, gdpr22automated}.:

\begin{enumerate}
\item Right to Access Data: The user should be able to access their collected personal data in a readily accessible format. For instance, allowing the users to review their preference tags as Google does\cite{google2021ads}. Therefore, the users can know what to expect to see in the advertising. This right aims to increase the Transparency of the platform.

\item Right to Opt out or in: The users should be able to opt in or opt out from any data collection/processing if it’s not directly involved in the core functionality of the platform. For instance, as a social audio streaming platform, advertising is not the main service. Therefore, the users should be able to opt out from the behavior tracking for the purpose of advertising. However, we recommend that the platform apply opt-in choice instead. In the same time, provide extraneous incentives, such as discounts to certain products, to encourage users to opt in to the behavior tracking. The opt-in choice empowers the users to have more control over the platform.

\item Right to Deletion: The user should be able to ask the audio platform to delete certain personal data. Users should generally be allowed to request the destruction of their data. Exceptions might include data that is required to complete a transaction or data the platform is legally required to retain. For example, the users can request to delete the credit card number collected by the financial services, if the users already acquired the product they bought. More detail will be discussed in the following paragraphs.

\item Right to Restrict Data Usage: This right grants the user to specify how certain data can be used under certain purposes. Ideally, if the purpose is not directly related to the platform’s core functionality, each purpose should be subject to opt-in consent, rather than opt-out. However, the platform tends to lack motivation to make opt-out controls easily accessible to users \cite{habib2020s, habib2019empirical, leon2012johnny}. On the other hand, if usage is subject to prior consent from the user (“opt-in”) the platform will work harder at making sure that the option is readily visible and can easily be used. In addition, “opt-in” significantly increases the chance that users are aware of data practices that are not core to its services, as they have to authorize these practices ahead of time. For example, the users should be allowed to restrict the use of their search history as one of the sources of improving target advertising. But keep the functionality for them to easily trace back to previous search items.

\item Right to Recitation: This right allows the user to review and correct erroneous information about them (e.g., erroneous information about their age or erroneous information about their history of posting inappropriate content. For instance, the users can correct their data of birth if they accidentally input the wrong information during registration.

\item Right to Object to Automated Decision Making: We also recommend that platforms explicitly notify users about their use of automated processing functionality such as data mining functionality and give users the ability to object to such processing. One example would be to allow users to object to the use of data mining to develop profiles about them. Such techniques have been known to sometimes have inherent bias and it is important for a platform to give its users control over whether or not they would like their data to be processed by these types of algorithms.

\end{enumerate}

Effectively supporting these Data Subject Rights requires that the platform does a good job at publicizing its support for these rights, making it easy for users to discover their existence and effectively take advantage of them, which is primarily a usability issue. It would be logical for the audio platform to provide access to functionality to exercise these rights in the dashboard that we already recommended earlier in this report. In addition, different data subject rights should also be accessible in context. For instance the ability to request the deletion of a particular stream should be available from the main screen for that stream as should controls to possibly restrict the possible collection and use of behavioral data  collected for a given stream. Furthermore, the submission of Data Subject Requests, such as deleting certain pieces of information, or correcting personal data, etc, are recommended to happen in the same place as well. The following paragraph will further discuss the details of the Privacy Dashboard. 

Links to the dashboard, including relevant privacy labels and controls, should also be provided in context, as already discussed above. It goes without saying that exercising one’s right should extend to third parties that might be storing or processing data. It would be unrealistic to expect users to identify every single third party and submit equivalent requests to each one of them. Instead, the platform has to take responsibility to support data subject rights locally but also for transmitting requests to relevant third parties handling the user’s data and for ensuring that these third parties honor these requests. For instance, if a user  requests the deletion of a piece of data shared with a third party  service, the platform has to pass the user’s request to the third party service and obtain confirmation from the third party that the data has indeed been deleted. In order to transfer or conduct the data subject request smoothly, the audio platform may need to agree on APIs to pass requests to third parties and receive confirmation that these requests are being processed. And the audio platform should consider this requirement before selecting any Third party services. Additionally, the user should not have to contact the third party directly. That is to say, the audio platform should not only provide a link or transfer the request to the third party and let the users directly interact with the third party. Instead, the platform should take responsibility for all the possible issues that might happen during the process.

\paragraph{Notification and Consent}

Only with proper notification, the users are able to make meaningful choices. The best practice of notice timing is to make the notice right before the data collection/processing happens \cite{schaub2015design} if it’s not disruptive to the users. The best timing depends on the usability, for instance, the notification of streaming recording could be shown before the users joining the streaming. It would be very disruptive if the notification jumped out every few seconds. Specifically indicates that piece of information will be shared with the third party service provider. Take the “credit card number” of a certain user as an example, the notification should at least include the following information.

\begin{enumerate}
\item The purpose/reason of data processing (“complete financial transaction”)
\item How the credit card number is collected (“input by user”)
\item When is the time the data is collected (“2021-10-19”)
\item Who can access the data for this purpose/reason? (“PayPal”)
\end{enumerate}

In general, users selecting to not opt-in to data practices that are not required for the platform to provide a service should not be penalized by being deprived of that service. For instance, a user should not be prevented from accessing a particular stream if he or she didn’t opt-in to the collection of behavioral data for marketing purposes. Business considerations may however dictate charging users differently based on how much data they are willing to have collected about them and for what purpose, as long as the process is transparent and users are given a clear choice.

A common practice of providing notice and choice is implementing a pop-out dashboard. In the dashboard, the audio platform indicates all the cookies and trackers running on the platform. Therefore, the users can evaluate all the services and choose to opt-out from them. However, considering the complexity of third party cookies, there might be hundreds of cookies, trackers and services running from dozens of providers simultaneously on the platform. It’s not accessible to present all information without categorizing. The best practice is to provide a dashboard with categories that enable the user to manage several services in a bulk. For instance, divide those services into the categories by the purpose like “financial”, “advertising”, “quality of life” and allow the users to opt out/in certain categories.

The default choice should consider the benefit of the user. A good default option should consider the user’s acknowledgement and requirement. Especially when the user has specified their requirement. For example, if the users already choose to not to share his/her personal information to a certain third party in the privacy dashboard, the default choice should disable the cookies, tracker and service that collect the personal data from a certain provider.

The platform should avoid using “Dark Pattern” when designing the interface for notice and consent. Dark Patterns refer to the design instance where the designer intentionally guides the user to do something not in the users’ best interest \cite{gray2018dark}. For instance, obfuscating the link to the privacy policy with lighter color, or unnecessarily arduous procedure to opt out from certain service, are all considered as examples of dark patterns.

However, with many notices and consent given out, the users might need a centralized page that aggregates all the consent ever given on the platform. Additionally, the users might want to change his/her choice after consideration. We recommend aggregating all the notice and consent in the form of “Privacy Dashboard”. The details will be discussed in the following paragraph.

\paragraph{Privacy Dashboard}

A Privacy Dashboard can serve as a one-stop shop screen, where users can review all data collected about them and exercise their data subject rights\cite{hoepman2014privacy}. Firstly, we often observe that many websites, such as YouTube, have privacy related information scattered all over the website, ranging from terminal agreement, creator policy, community guidelines, third party policy, etc. As a result, the information provided by the platform might be inconsistent. Also, the users might be confused to find the data practice disclosure associated with certain features as well. The best approach is to provide a summary for all the data practice disclosures. The summary is categorized in different themes, for instance, financial, tracking, etc. To further help the users to better understand how the platform applies the disclosures, we recommend visualizing them. The “Privacy Nutrition Label” applied by Apple provides a good example of visualizing the privacy policies \cite{kelley2009nutrition, apple2021aday}.

The Privacy Dashboard also allows users to review their personal information. Notably, the personal information is not only the data directly collected from the users. It also includes the data derived from the collected data, for instance, the advertising preference based on search result \cite{google2021ads}.  Furthermore, the dashboard should also include the collected information of third parties as well.

The objective is to make it as easy as possible for a user to review all data collected about them and exercise their rights as they pertain to this data, all from a single location. To further achieve this goal, we recommend that the audio platform aggregate the information by the collected purpose. The audio platform should avoid grouping the information items in by the collection parties (internal teams, third parties), or collection method (tracker, cookies, ML model, etc). People tend to be confused by those categories \cite{leon2012johnny}. It would be easier for users to navigate those items by organizing content along several different dimensions, such as timing of collection, type of content, public or private accessible, etc.

The other recommendation is for the audio platform to provide access to Data Subject Requests from the same Privacy Dashboard. Data Subject Requests should include requests to change, delete or access one’s. The requests directly map to the Data Subject Rights mentioned in the previous paragraph.

However, if the system is not developed to deal with data requests in the first place, it would be extremely difficult to add the mechanism afterward. The privacy requirement should be considered in the very first phases of the software development of ideation and definition. In the design phase of the software development, we recommend adopting software more flexible design patterns such as RESTful API. For existing systems to address data subject requests, the best practice is to compose a standard procedure to aggregate required information that is assigned to a certain team. This may require the audio platform to re-architect some of its functionality.

After a user has submitted a request, for the sake of Transparency, the platform should provide users with feedback on the processing of his or her request, including an estimate of when the request will have been fully processed. Such feedback should include  expected completion date, a timeline that shows the current status of the data item, and possibly a log summarizing steps that have already been completed, etc. With this information, not only can the user have a clear understanding. It also provides an accessible interface for the responsible team to monitor progress of the request. 

As mentioned in the previous paragraph, the privacy dashboard could also be the place to record all the consent ever done by the users. The record shows where and when the user gives the consent to allow them to collect/access certain information. For instance, the record of credit card number may be presented as follows:
\begin{itemize}
\item Event: Share credit card number to PayPal
\item Data: Credit card number 1234-1234-1234-1234
\item Time of consent: 2021-01-01
\item Place of consent: Making transaction \#05664 (may be a link)
\item Consent: Yes
\item Involved Parties: The Platform, PayPal (may be a link)
\item Original Notification: link to original notice text
\end{itemize}

In order to improve the Individual Participation principle. The dashboard should also allow the user to retrieve the consent. At least, allow the user to submit the request to the involved parties to alter the consent.

\subsection{Best Practice of Internal Data Usage}

An internal policy of data usage that concerns the user privacy is the basis to implement the best practices. In order to make every employee acquire the common understanding of the policy, we recommend that the platform provide training procedures and evaluation systems. The training session should try to convey only the necessary information. For instance, the employees of the site reliability team do not have to go through the training session about financial transactions. The excessive content would overwhelm the employee and diminish the efficiency of the training \cite{kumaraguru2010teaching}.

A centralized database schema that indicates the information flow across the audio platform would greatly improve the usability to audit the information flow. Serve as a central-hub of personal information usage. To implement the feature, the database schema based on the relational database is recommended, since the relational database can reduce the complexity of many-to-many mapping of the information flows that involve different departments in the audio platform, users, and third parties.

In order to exercise the Purpose Specification Principle, before designing any new feature, the potential data usage and sharing should be evaluated. The consideration of privacy should be included in the very first place of software development. That is to say, before developing any functionality that involves accessing certain personal information, the platform should first evaluate the necessity of using the information. If necessary, apply required modifications to the data. We recommend that the audio platform apply the strategies mentioned in Hoepman’s work of \textit{Privacy Design Strategies}:

\begin{itemize}
\item Minimise: Only allow the specific parties (could be persons or automated programs) to access a limited amount of data. For instance, in order to complete a purchase, only select the information that is associated with this purchase and share it to the service, eg. credit card number, authentication token. But not allow the service to access other information like purchased records, product reviews, etc.
\item Abstract: Obfuscated the personal information in the high level. For example, provide aggregate statistics instead of specific records of personal information. Providing mocking information for the developers instead of working on the real user data.
\item Hide: The strategy aims to make personal information unlinkable to the specific user. The transaction of information should always be in the encrypted channel like HTTPs. Setup access control policy that ensures only the authorized parties can use it. Mix the source of personal information by processing them in bulk.
\end{itemize}

To further reduce the invalid data sharing and usage, we recommend the audio platform to apply Access Control policy. Access Control is a series of policies that provide a group of user privileges to certain files, or data. It can be applied to constrain how the programs execute on behalf of the user \cite{sandhu1994access}. To address the complex many-to-many issue of information flow, we recommend applying a Role-based access control (RBAC) scheme. The scheme is designed to reduce the complexity and cost of information administration. The platform first identifies the different “roles'' across the platform. A role can be created according to the team's functionality, or assigned according to the employee's position. A single employee might hold several roles. To make any change of certain employee’s accessibility, the administrator simply removes old roles and entitles new roles to that employee \cite{ferraiolo2003role}. Each role consists of several “privileges” across the data. The privileges are composed heavily based on the application. For instance, the privileges in SQL are “INSERT'', “DELETE” or “TRIGGER”, etc. Creating a new role should be reviewed more cautiously based on the Use Limitation principle. The platform should give least privileges to the roles. If any certain employee, or programs are working on a very specific requirement of privileges, instead of creating a new role, the platform should consider assigning existing roles to that instance.

\subsection{Third Party Data Usage}

Most, if not all, social audio streaming platforms rely on third party services for a variety of different purposes, including cookies, analytics, tracker, content, advertising, social network integration, payment, etc. Fair Information Practice Principles apply to these interactions as well. The social audio platforms should try to extend their best practices to their Third Parties partners as well. We recommend the audio platform to develop a concrete and universal third party policy. The policy should at least cover the following topics:

\begin{itemize}

\item Third party is able to handle transferred Data Subject Requests and establish a formal channel to contact with the social audio platform.

\item Third party provides sufficient information to both the social audio platform and the users about how the services collect and use personal information

\item Third party should not share or sell certain types of personal information without user’s consent

\item Third party is willing to provide practical choice of opt out-in of their services

\item Third party commits to comply with all applicable privacy regulations and to process data in a manner consistent with the audio platform’s policy. 

\item Third party has concrete privacy policies and publicly publishes them

\item All of the requests, consents, and choices can be done without the interactions between the users and the third party. For instance, the users only have to change the privacy settings in the social audio streaming platform to opt-in behavior tracking by the third party. The users do not have to submit another request to the third party about these requests.

\end{itemize}

In order to ensure third parties comply with the policy, we recommend the social audio platform to compose a checklist that covers the mentioned topics. The checklist should also include the legal requirements and business considerations as well.
\section{Best Practices for Data Destruction}

Data destruction is the final stage of the data life-cycle where the data has either fully served its purpose or the user has requested that it be deleted. The procedure for submitting the  removal/destruction request should be easy and can be carried out by any users. The social audio platform should not only consider where and how to carry out the destruction procedure. Additionally, the social audio platform should demonstrate the progress of the data destruction to the users as well.

\subsection{Retention Period and Automated Data Destruction}

Retention Period indicates how long the social audio platform keeps the data in the storage space. The retention period of personal information should be kept to a strict minimum \cite{cavoukian2009privacy}. That is to say, once the original purpose is served, the platform should delete the data. Social audio platforms should compose a retention schedule for every collected personal information item. The retention period should also be noticed to the users as well. We recommend that the platform notice the information of retention periods when the data collection happens. The notice includes, retention period and the specified reason.

If the platform has plans to archive data beyond the retention period, the platform should give explicit notice and consent to the users. For example, send an email to notify the users of archiving customer service cases for internal training. In this notification, the platform indicates the reason and applied methods to obfuscate the data. The issue regarding archiving data will be discussed in the other section.

Furthermore, if the data contains multiple columns, the platform should consider dropping certain columns that are not required for the future usage. For instance, in order to train a prediction model for item purchasing, the platform requires the purchase history. The purchase history might contain multiple columns such as user name, credit card number, timestamps, quantity of items, etc. However, the platform does not require the credit card number and user name to train the model. All they need is the time and quantity of certain items being purchased. Therefore, the credit card information should be removed after the transaction is done.

Data deletion should systematically take place when the stated retention time has elapsed. Data deletion could be done manually or automatically. An automatic process is generally recommended because it is more efficient and because it also guarantees that deletion will be done in a systematic manner.

Data can become a liability if it is retained for longer than needed, as it remains subject to possible data breaches. In addition, not deleting data before the expiration of that data’s retention period exposes a company to regulatory actions\cite{geambasu2009vanish}.

However, there are several challenges to implementing automated data destruction  mechanisms for the social audio platform. The nature of a distributed system is that it is very hard to confirm the status of other machines. The synchronization issues of deleting all data at the same time is also very challenging to handle \cite{zeng2010safevanish}. Fortunately, most commercial cloud services provide automated data destruction functions that are highly usable. For instance, in AWS S3 storage service, there is a "Lifecycle configuration element" for every object \cite{s3life}. This configuration can either be set to \textit{Expiration} that deletes the object or \textit{Transition} that transits the object to specified storage spaces. The \textit{Expiration} function is ideal for implementing the retention schedule. And the \textit{Transition} function is highly recommended for archiving data.

\subsection{Enforcing Data Deletion Request}

\subsubsection{Usability of Data Deletion Request}

A clear, intuitive UI design is the first step of the best practice to exercise the data subject request. Without obvious UI components, the users might not be aware that such functionality even exists \cite{habib2020s}. The following list shows some examples of obvious UI components that might help the user to be aware of their data subject rights.

\begin{itemize}
    \item An obvious button in account setting page with high contrast color with the word of “DELETE MY DATA” inside
    \item An icon of garbage bin or red cross marker by the data item in the personal information disclosure page
    \item A dedicated page to submit data deletion requests that can be entered via a highlighted link in privacy settings and/or policy page, such as “DELETE MY DATA”
    \item A red word of “DELETE MY DATA” in side menu of the APP that guide the user to the submission page
\end{itemize}

The platform should evaluate the usability of any UI changes before pushing it online. Usability here should be interpreted as encompassing multiple considerations as discussed by Feng et al. \cite{feng2021design}. The best approach for evaluating usability for new UI components is going through semi-instruction interviews \cite{habib2020s}. Semi-instruction interview is the combination of task observation and interview. The interviewer only asks a few predetermined questions. The user might be asked to find out and submit a data deletion request for a certain item\cite{habib2021evaluating}. The interviewer then derives the follow ups questions by observing the tasks. For instance, the follow up questions such as “Why are you confused when choosing the category”, if the interviewee showed hesitation when choosing the category of request. It often only requires 10-20 people to conduct these types of studies \cite{lazar2017research}.

\subsubsection{Recommended Backend Design}

In the perspective of backend system design, it could save a great amount of time and effort if the social audio platform applied the RESTful API design principle\cite{masse2011rest}. RESTful API provides a uniform interface and centers around resources. Therefore, for every collected data item, there should be a matching API to delete it. There's no need to develop an extra layer of API to delete certain data. However, this requires a proper normalization to database schema \cite{beeri1989sophisticate}. Database normalization aims to eliminate the redundant data entries across all of the database system. Hence, there is no need to go through all the databases to delete certain data items. It also reduces the risk of accidentally leaving the data item behind somewhere in the database.

\subsubsection{Obligation to Third Parties}

The social audio platform should be able to delete the information collected and/or storage by third parties as well. In order to exercise the Accountability Principle, the platform has an obligation to work with their third party providers to comply with the best practices. The implementation should be more than merely redirecting the users to submit yet another request to the third parties.

\subsubsection{Demonstrate the Progress}

In order to improve the transparency of social audio platforms, we recommend the platform to demonstrate the progress of data destruction. The best practice here is reuse the common UI components that can instantly evoke the familiarity of the users. Therefore, it reduces the extraneous time and effort to understand the conceptual functionality of the component \cite{tognazzini2003first}. Progress bar is a common component to show the progress of long tasks. Most prevalent front-end frameworks provide the default components to implement progress bar \cite{bootstrapProgress}. And users have a strong preference for progress indicators during long tasks \cite{myers1985importance}. We also recommend that the platform includes the progress bar in the same page of submission to create a one-stop solution for similar issues, such as the Privacy Dashboard mentioned in the previous section.

\subsection{Archiving Data}

There are some scenarios where the social audio platform can archive collected information beyond the retention period. However, we recommend that the data should be modified to minimize potential privacy and security risks. At a high level, there are two major strategies to protecting archived personal information: Hide and Abstraction \cite{hoepman2014privacy}.

To hide the data means to make the data unlikeable to the original source of users. Some obvious methods include, encrypt the data, store the data in a separate database and reduce the possibility of mapping the data to the users. In addition, we recommend to anonymise the data. Anonymization relies on a collection of methods to ensure that even if a data breach occurs, the risks that an individual be re-identified are minimized \cite{pfitzmann2001anonymity}. However, anoymization is not simply using the pseudonyms to store user data. The platforms need concern about the risks of reidentification associated with opportunities to combine data collected by the platform with other sources of data such as public datasets or possible data available from third parties. Anonymization is more than just arbitrarily obfuscating data. There are also tradeoffs between the utility of sanitized datasets and the privacy guarantees they offer. These tradeoffs can be quantified using metrics such as k-anonymity, l-diversity, and t-closeness \cite{pfitzmann2001anonymity, narayanan2006break}.

Abstraction indicates to limit as much as possible the detail during the process of personal information. There are many applications that do not require storing individual data. For instance, we do not always need to record the purchase history of every individual to conduct marketing analysis \cite{louviere1983design}. Instead, it is often possible to identify important trends using aggregate data such as, “how many users aged between 20 and 30 who purchased product A within the last 3 months also purchased product B?”

\subsection{Legal Basis}

\subsubsection{APPI}

The right of data destruction is required by APPI when the data is no longer necessary for the platform to operate. Notably, APPI does require that data collected by third parties should be deleted as well.

\subsubsection{CPRA}

Many of the above best practice recommendations are aligned with requirements associated with CPRA. There should be a clear notice of the right to request data deletion  in an obvious way. CPRA also highlights that the platform has the obligation to their third party providers. Under no circumstances, the consumers have to visit the third parties by themselves. Furthermore, the platform must track the progress on the third party’s side as well. There are few exemptions that the platform could hold certain data longer than expected. The exemption purposes include: the data is required to complete a transaction; improving safety; debugging; internal uses; legal obligations; freedom of expression.
\section{Moderating Guidelines}

Content moderation is essential to maintaining user trust and safety. As such it can be viewed as an extension of privacy protection. Throughout this section, we use the common definition of \textit{Safety} as \textit{freedom from unacceptable risk} \cite{hofmann2011error}. There are multiple categories of online risks such as harassment, inappropriate content, hate speech, etc. In many situations, these risks are directly related to privacy considerations.

For instance, ‘doxing’ is the act of revealing someone’s private information publicly. One example of doxing is to disclose a rival’s home address online to allow followers to send unwelcome mail, trespass, vandalize, threaten, etc. This is an evident security risk that falls under the category of harassment. It is also an evident privacy risk in that a home address is personal information that the data subject should have control over.

A common definition of privacy involves the right to be ‘let alone’, free from undue harm from other people or entities \cite{parker1973definition}. When a user creates audio content that discloses private information about someone, spreads false rumors about that person, or entices others to do harm to that person, that user infringes on that person’s privacy. On the other hand, when a platform moderator unreasonably removes a user’s content they are infringing on that creator’s freedom of expression, which is another facet of privacy. We discuss possible best practice recommendations aimed at balancing these different considerations and at doing so in a fair, transparent and cost-effective manner.

We acknowledge that content moderation is a challenging problem and one that does not yet admit well established best practice recommendations. Our analysis and recommendations are organized around a multi-layered approach to content moderation similar to the one discussed by Twitch in its “Transparency Report 2020” \cite{twitch2021transparency}. Just like with Twitch, the majority of content available on social audio streaming platforms is ephemeral in nature, requiring an approach that is somewhat different from that of a platform that primarily hosts pre-recorded content. Ensuring safety in this context requires an approach that combines efforts of the audio streaming platform itself (though a combination of tools and personnel) and input from community members/users. As with Twitch we distinguish between four different levels of safety:

\begin{enumerate}

\item Community Guidelines: Community guidelines establish baselines. They outline general objectives and principles associated with a given community of users. Guidelines include general principles that every user should aim to follow as well as more specific rules that may mandate and/or prohibit some behaviors, effectively establishing a code of conduct. As such community guidelines help set expectations among users and may also be used by prospective users to determine whether a given community conforms to their values and/or expectations. This is further discussed under “Defining Violations to Community Guidelines”, and “Identifying Violations to Community Guidelines” in subsections below.

\item Platform Level: This level aims to enforce community guidelines through a combination of automated tools (e.g., machine learning tools, word filters), user reporting and manual content moderation by platform personnel in charge of reviewing results produced by automated tools and of adjudicating reports from the user community.

\item Creator Level: Creators should also be given functionality that enables them to specify and enforce channel-specific guidelines that are somewhat stricter than those adopted by the platform as a whole (e.g.,  channels dedicated to children or other groups of users who generally desire or require stricter standards). These stricter guidelines should not be allowed to conflict with the general principles and guidelines set for the platform as a whole.

\item Audience Level: Individual users should also be given functionality that enables them to further customize the content they see. This might include support for different sets of filters (e.g. filtering out topics or content that a particular user might find offensive even if this content does not contravene the platform’s general guidelines)

\end{enumerate}

\subsection{Defining Violations of Community Guidelines}
Community Guidelines are a set of principles and rules that define safety guardrails that apply to all users and their activities on the platform Each social audio streaming platforms may want to have its own specific guidelines designed to accommodate the particular sensitivities of the users it caters to and the various laws under which it operates – with possibly different guidelines applying under different jurisdictions. Guidelines will reflect the purpose of the platform, potential dangers associated with content that might be shared on the platform, legal obligations, philosophies on social responsibility, etc. In this section, we discuss community guidelines from the perspective of privacy. We discuss best practices for defining and identifying violations of such guidelines.

To ensure transparency and fairness, the social audio streaming platform will want to clearly identify categories of safety and privacy violations that it deems unacceptable. Such definitions typically benefit from being illustrated with concrete examples. Common categories of violations that apply to a social audio streaming platform include Inappropriate Content, Impersonation, Bullying, Hate Speech, Suicide/Self Harm, Suspected Minor, and Violence Threats.

We also recommend including an additional category that could be referred to as ``third person disclosures” to capture scenarios where a user divulges personal information about a third person- whether this person is on the platform or not. This specific violation touches on one of the main challenges involved with processing user-generated content. The contents of user-generated content can be extremely sensitive and possibly factually false. A platform should consider including functionality that allows someone to request that false information spread about them be taken down. The platform might want to also support functionality that signals to users that some content is being contested. Another challenge is when a content creator divulges non-public, sensitive information about a third person even if that content is factually correct (e.g., someone’s marital problems, sexual orientation, problems with alcohol, etc.). Minimally, the platform should have a code of conduct that warns users against divulging such content. It also has a general obligation to educate its users and encourage them to carefully think about the content they publish. The platform could also offer functionality that helps people document incidents and request that some content be taken down. But it might want to distance itself from actually having to directly adjudicate some of these scenarios, and instead it might defer to the legal system.

Allowing users to potentially divulge a third person's personal data creates many moderation and organizational challenges. Would the data controller be required to notify the third person that they are now collecting personal or sensitive information about them? How would the data controller contact the third person if they are not also a user of the system? These issues are topics that the privacy and moderation teams will want to carefully discuss with management and legal.

Harassment is a particular type of violation that draws special attention. Harassment is sometimes clear cut but can also sometimes be subjective with content being viewed as being protected by free speech by those who create it while being viewed as offensive, intimidating and embarrassing by those who are the subject of that content. Past literature shows that victims of harassment are often likely to self censor and not reported as such \cite{ong2021online} \cite{chadha2020women}. Accordingly, an important part of providing people with the safety they need involves providing an environment where possible victims of harassment feel safe and comfortable enough to report it.

In addition to clearly defining categories of violations and a clear code of conduct, an audio streaming platform should ensure that members of its community understand the rules and expectations, using enough specific examples to at least cover some of the most likely situations.
Using examples is an effective way of explaining violations to users \cite{seering2017shaping}.While requiring users to take some training prior to being able to use the platform seems unrealistic, as it would likely deter many from joining the platform, nudges could be used to encourage users to reflect on the content they generate and consult guidelines, including examples, when unsure. The nudges could also direct users to gamified interactive training content. This practice has been shown to be quite effective in other security and privacy contexts (e.g. \cite{sheng2007anti, acquisti2017nudges}). 

In the coming sections, we focus on best practices for reporting and enforcing third person disclosures and harassment.

\subsection{Identifying Violations to Community Guidelines}

Traditional moderation tools are not effective when applied to audio data, especially ephemeral audio data \cite{jiang2019moderation}. Here we discuss one approach for identifying violations in a social audio streaming  platform: trusted user reporting.

The benefit of allowing trusted users to report content is clear, it enables organizations to identify violations quickly and address high-profile scenarios before they develop into public relations scandals. Many social platforms employ some form of user reporting - Facebook and Twitter allow users to report content, and Youtube has a Trusted Flagger program \cite{youtubetrustedflagger}.  

However, there are also many challenges involved including motivating a user to report, preventing abuse of this functionality, and ensuring that reports are actually useful in prioritizing a moderator's time. 

In order to motivate a user to report and prevent abuse of this functionality, we propose an incentive system. In this incentive system, a new user is given certain neutral ``score". All reported content is reviewed by the moderating team and acted upon if there is a violation, or dismissed if there is no violation. Whenever a user appropriately reports content, meaning the moderating team determined there was a violation, the user's ``score" is increased by a set amount. Likewise, if the user reports content that did not contain a violation, the user's score is decreased by a set amount. Defining a scale for the score is an organizational decision. It could be numeric (1 - 5), textual (not trusted, neutral, trusted), or even icon based.

There should be thresholds for determining when a user is considered ``trusted" or ``not trusted". Deciding how to transparent to be about the threshold for defining a trusted or not trusted user is also an organizational decision. Lack of transparency could lead to claims that the decisions are arbitrary while complete transparency could enable users to abuse the tool. 

For the incentive component of this system, users who receive a trusted status could be given recognition in the form of a badge next to their name, and/or a simple perk. The badge functions as a social incentive whereas the perk functions as a monetary incentive, and they could appeal to different users. Using Spoon Radio's platform as an example, trusted users could be given a free sticker as a perk. 

Users might abuse reporting functionality by reporting content they simply don't like, or reporting a large amount of content to overwhelm moderators or the underlying infrastructure. Preventing users who are not trusted from reporting limits the scale of these abuses. 
Other best practices include making icons used to report content clear to users and easily accessible. Best practices recommend conducting a usability study to ensure icons are accurately conveying what the social audio streaming platform intends them to \cite{habib2021toggles}. User's should be given representative examples of violations, both in the interactive format recommended above and with more detail if desired. 

Lastly, we discuss a recommendation intended to make user reporting useful to moderators. Standard user reporting merely flags content for review, more useful reporting would help prioritize this content. We suggest a layered system of reporting that remains short so as to not burden users and discourage them, but asks key questions that are used for prioritizing. In this layered system, the user first selects the category of violation to be reported and is then asked one or two key questions. 

Using harassment as an example, the user might be asked:
\begin{itemize}
    \item Is a specific individual being targeted? Please include their handle if they are a user. If you don't know the handle you can still report this post. 
    \item Is the creator calling for harm to an individual? (i.e. disseminating their home address without their consent and encouraging users to show up or send them mail)
\end{itemize}

It should be clear that these questions are not mandatory to report the post, and a reporting user should be allowed to answer ``I'm not sure" and report the post anyway. The answer options should be yes/ no/ I'm not sure, and there should be a text box available for explanation. 

\subsection{Platform-Level Safety}

The platform moderator is a group of employees with sufficient training to address the violation events across the platform. In order to act in a fair, consistent and transparent manner, the platform should ensure that its moderators are adequately trained in reviewing content. Such training should include training in reviewing content, making decisions and documenting these decisions. Their training has to cover a range of relevant knowledge, including linguistic knowledge (e.g., language used by relevant community of users), social-political knowledge, including sufficient background about past and current events, as well cultural and legal knowledge. A global audio streaming platform will likely want to have multiple teams of people with different teams responsible for content in different languages and different regions of the world.

In general, platform enforcement should be fair, consistent and transparent. This means that when the platform takes an action, such as taking down some content or deciding not to take down some content, it should make sure to justify its actions. The platform should also maintain a database with all these cases it has handled, their history, context and resolution and allow its moderator team to easily access the content of the database to help inform their decisions as they review new incidents. Particular egregious behaviors from some users could lead the platform to not just take content down but to possibly suspend some users for some period of time or even permanently.

The primary task of moderators is to review the content reported by users or flagged by automated tools. Seldom should the moderators have to proactively review content, as this is just impractical. We suggest the moderators follow the best practices shown in the following flow charts \cite{veglis2014moderation}.

\begin{figure}[htbp]
    \centerline{\includegraphics[width=\columnwidth]{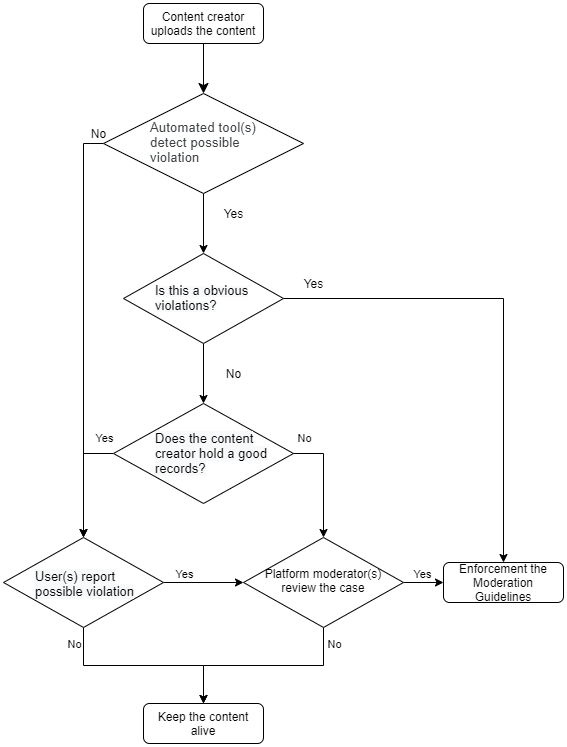}}
    \caption{Overall flow-chart for content moderation.}
    \label{moderation-flow}
\end{figure}

The big difference between the ephemeral streaming data is that there’s no time window for the moderators or automated tools to review the content right before its publication. Since the upload and publication happens at the same time. Another challenge is that there are no robust automated tools for dealing with real-time audio data. Yet the development of such automated tools could greatly improve the quality, efficiency, and consistency of the review of the review process.

Obviously automated tools will always have their limitations and the platform will want to be transparent about its use of such tools and explain how results from automated tools are supplemented with manual analysis by professional content moderators. Being transparent about content flagged by automated tools pending a manual review would help a platform come across as being efficient while also acknowledging the inherent limit of automated tools.

\subsection{Creator/Channel Level Safety}

In order to empower the content creators to have more influence with community building, we recommend that the platform allows the content creators to impose some additional restrictions on acceptable content and behavior of their channel. The channel owner should carefully document these rules as well. The platform moderators might want to review the rules to make sure they are consistent with those of the platform as a whole.

The channel moderators are a team of users that help the channel owner to enforce the channel's guideline. Notably, platform moderators should have the ability to overrule channel moderators, the latter typically being platform users rather than platform employees. This might require the implementation of workflows to help establish the review process – between the channel moderator, users who report inappropriate content, possibly mechanisms for people whose content is taken down to appeal decisions with the platform, etc.

\subsection{Listener Level Safety}

At this level, the goal is to empower users to control the content that shows in their feeds. This  aligns with the Individual Participation principle. Firstly, allow the user to block certain content. The users can block the content that is generated by certain creators or hashtags. This might involve using a set of predefined filters made available to each user by the platform.
We recommend applying the content rating system like how movie, and video games have\cite{youtube2021Content}. There are the ratings of “Matured” or “General” content. The “Matured” content is only available to adult audiences.

Another recommended implementation is the content warning system. Although the audience is an adult, there might be certain types of content that the audience does not wish to be exposed to. The content-creator or the platform could  label content with specific topics and show the warning right before playing the content. The warning could indicate what might include in the content, such as sexual, violence, horror, etc. Accordingly a user would be able to configure settings that would ensure he or she does not get to interact with this content.

\section{Summary}
Throughout this report, we compiled a list of best privacy practice recommendations for a social audio streaming platform. We organized these best practice recommendations around the data life cycle, including an additional discussion about content moderation. Collection and processing of unstructured, user-generated content gives rise to a number of unique challenges. Moderating unstructured content requires more resources and innovation than moderating structured content. Since user-generated content can include information about people who are not the content creators or even users of the platform, issues of deletion, access, and correction are also more complex.

We recommend that audio streaming platforms provide dashboards that allow users to readily access data collected about them and allow them to exercise a rich set of data subject rights, including correcting their data, requesting that it be deleted, but also restricting the way it is processed (e.g., automated processing) and used, including sharing with third parties. We recommend that this dashboard rely on succinct data practice disclosures in the form of privacy nutrition labels. 

We also recommend that similar notices and controls be made available throughout the platform to ensure that users are aware of what data is collected about them, how that data is used and how they can go about exercising their rights about the collection and use of that data. We strongly recommend adopting a framework in which all third parties are required to honor the same privacy commitments made by the platform and provide APIs to ensure that data subject rights extend to data shared with third parties. Our recommendations extend to the adoption of fair, transparent, consistent and efficient content moderation practices aimed at protecting the safety of the platform’s users while respecting their freedom of expression. 

Our recommendation in this area rely on combining the use of automated tool with the processing of reports from the user community and the support of a team of well trained moderators responsible for making final decisions. Our recommendations include an incentive-based user-reporting system. This incentive-based system will appeal to different users by providing social and financial incentives, as well as helping mitigate abuses of user-reporting by eliminating problematic users. We  recommend layered user-reporting so that a report not only flags content, but also helps moderators prioritize especially problematic or dangerous content.

We believe that by following the best practice recommendations summarized in this report, an audio streaming platform such as Spoon Radio will be well prepared for future changes in privacy regulations and for continued expansion in other markets.
We hope that best practice recommendations in this report will prove useful to Spoon Radio as well as other social audio streaming platforms. There are various components of this report that will require further analysis and monitoring, such as content moderation and data minimization issues including retention policies.

\section*{Acknowledgment}
We would like to thank the Carnegie Mellon Privacy Engineering Master's program for providing us with the education and tools necessary to complete this report. We would like to thank Professor Norman Sadeh for his continuous support and feedback. Finally, we would like to thank those employees at Spoon Radio who spent time interacting with us and sharing crucial insights into the audio streaming industry, Ye In Kim, Dong Hyuk Shin, Fernando Pizarro, and Spoon Radio Management.

\printbibliography
 
\end{document}